\documentclass[prb,preprintnumbers,amsmath,amssymb,floatfix]{revtex4}

\usepackage{graphicx}
\usepackage{subfigure}
\usepackage{dcolumn}

\begin{document}

\title{Quantum entanglement degrees amplifier}
\author{Xiang-Yao Wu$^{a}$ \footnote{E-mail: wuxy2066@163.com},
 Ji Ma$^{a}$, Xiao-Jing Liu$^{a}$\\ Yu Liang$^{a}$, Xiang-Dong Meng$^{a}$, Hong Li$^{a}$ and Si-Qi Zhang$^{a}$}
 \affiliation{a. Institute of Physics, Jilin Normal
University, Siping 136000 China}
%%%%%%%

\begin{abstract}
The quantum entangled degrees of entangled states become smaller
with the transmission distance increasing, how to keep the purity
of quantum entangled states is the puzzle in quantum
communication. In the paper, we have designed a new type
entanglement degrees amplifier by one-dimensional photonic
crystal, which is similar as the relay station of classical
electromagnetic communication. We find when the entangled states
of two-photon and three-photon pass through photonic crystal,
their entanglement degrees can be magnified, which make the
entanglement states can be long range propagation and the quantum
communication can be really realized.

\vskip 10pt


\end{abstract}

\vskip 10pt \maketitle {\bf 1. Introduction} \vskip 10pt

Quantum entanglement is a unique phenomenon and its distribution
over a long distance is of vital importance in quantum
information. Many quantum processes require entanglement [1-3].
Quantum entanglement is the key source in current quantum
information processing [4]. Most quantum communication protocols
such as quantum key distribution [5], teleportation [6], quantum
secret sharing [7], quantum secure direct communication [8], and
quantum state sharing all need the entanglement to set up the
quantum channel [9].

Currently, the most important problem for single-photon and
two-photon entanglement may be the quantum repeater protocol in
long distance quantum communication [10, 11]. We know the quantum
entangled degree should be decreased and even approach zero, the
quantum entanglement of two and three photon shall disappear. All
the quantum information processes, e.g., quantum communication,
quantum computation and so on can not proceed. The attenuation of
quantum entangled degree is unavoidable, we can only make the
quantum entangled degree magnify in the entangled states
transmission process, which can be realized one-dimensional
photonic crystal.

Photonic crystals (PCs) are artificial materials with periodic
variations in refractive index that are designed to affect the
propagation of light [12-15]. An important feature of the PCs is
that there are allowed and forbidden ranges of frequencies at
which light propagates in the direction of index periodicity. Due
to the forbidden frequency range, known as photonic band gap (PBG)
[16-18], which forbids the radiation propagation in a specific
range of frequencies. The existence of PBGs will lead to many
interesting phenomena. In the past ten years has been developed an
intensive effort to study and micro-fabricate PBG materials in
one, two or three dimensions, e.g., modification of spontaneous
emission [19-22] and photon localization [23-27].

Due to various unavoidable environmental noise, the quantum
entangled degrees of entangled states become smaller with the
transmission distance increasing. How to keep the purity of
quantum entangled states is the puzzle in quantum communication.
In the paper, we have designed a new type entanglement degrees
amplifier by one-dimensional photonic crystal, which is similar as
the relay station of classical electromagnetic communication. We
find when the entangled states of two-photon and three-photon pass
through photonic crystal, their entanglement degrees can be
magnified, which make the entanglement states can be long range
propagation and the quantum communication can be really realized.

\vskip 8pt
 {\bf 2. Transfer matrix and transmissivity of one-dimensional
photonic crystal} \vskip 8pt

For one-dimensional conventional PCs, the calculations are
performed using the transfer matrix method [18], which is the most
effective technique to analyze the transmission properties of PCs.
For the medium layer $i$, the transfer matrices $M_i$ is given by
[18]:
\begin{eqnarray}
M_{i}=\left(%
\begin{array}{cc}
 \cos\delta_{i} & -i\sin\delta_{i}/\eta_{i} \\
 -i\eta_{i}sin\delta_{i}
 & \cos\delta_{i}\\
\end{array}%
\right),
\end{eqnarray}
where $\delta_{i}=\frac{\omega}{c} n_{i} d_i cos\theta_i$, $c$ is
speed of light in vacuum, $\theta_i$ is the ray angle inside the
layer $i$ with refractive index $n_i=\sqrt{\varepsilon_i \mu_i}$,
$cos\theta_i=\sqrt{1-(n^2_0sin^2\theta_0/n^2_i)}$, for the $TE$
wave $\eta_i=\sqrt{\varepsilon_i/\mu_i} \cdot cos\theta_i$, for
the $TM$ wave $\eta_i=\sqrt{\varepsilon_i/\mu_i} /cos\theta_i$,,
in which $n_0$ is the refractive index of the environment wherein
the incidence wave tends to enter the structure, and $\theta_0$ is
the incident angle.

The total transfer matrix $M$ for an $N$ period structure is given
by:
\begin{eqnarray}
\left(%
\begin{array}{c}
  E_{1} \\
  H_{1} \\
\end{array}%
\right)&=&M_{B}M_{A}M_{B}M_{A}\cdot\cdot\cdot M_{B}M_{A}\left(%
\begin{array}{c}
  E_{N+1} \\
  H_{N+1} \\
\end{array}%
\right)
\nonumber\\&=&M\left(%
\begin{array}{c}
  E_{N+1} \\
  H_{N+1} \\
\end{array}%
\right)=\left(%
\begin{array}{c c}
  A &  B \\
 C &  D \\
\end{array}%
\right)
 \left(%
\begin{array}{c}
  E_{N+1} \\
  H_{N+1} \\
\end{array}%
\right),
\end{eqnarray}
where
\begin{eqnarray}
M=\left(%
\begin{array}{c c}
  A &  B \\
 C &  D \\
\end{array}%
\right),
\end{eqnarray}
with the total transfer matrix $M$, we can obtain the
transmissivity $T$, it is
\begin{eqnarray}
T=|t|^2=|\frac{E_{tN+1}}{E_{i1}}|^2=|\frac{2\eta_{0}}{A\eta_{0}+B\eta_{0}\eta_{N+1}+C+D\eta_{N+1}}|^2.
\end{eqnarray}
Where $E_{tN+1}$ and $E_{i1}$ are the electric field intensity of
output and input,
$\eta_{0}=\eta_{N+1}=\sqrt{\frac{\varepsilon_0}{\mu_0}} \cdot
\cos\theta_0$ for $TE$ wave,
$\eta_{0}=\eta_{N+1}=\sqrt{\frac{\varepsilon_0}{\mu_0}} /
\cos\theta_0$ for $TM$ wave. By the Eqs. (1) and (4), we can
calculate the transmissivity of one-dimensional photonic crystal
for $TE$ and $TM$ wave.

\vskip 10pt {\bf 3. The two-photon and three-photon polarization
entangled states and quantum entanglement degree} \vskip 10pt

The two-photon polarization entangled state
\begin{eqnarray}
|\psi\rangle=\frac{1}{\sqrt{2}}(|HH\rangle+|VV\rangle),
\end{eqnarray}
three-photon polarization entangled GHZ state
\begin{eqnarray}
|GHZ\rangle=\frac{1}{\sqrt{2}}(|HHH\rangle+|VVV\rangle).
\end{eqnarray}
Here the states $|H\rangle$ and $|V\rangle$ represent the
horizontal (TM) and the vertical (TE) polarized single photon
states.

Due to various unavoidable environmental noise, the entanglement
of quantum entangled states become worse and worse with the
transmission distance increasing. So, how to keep the purity of
quantum entangled states is the puzzle of quantum communication.
The Eqs. (5) and (6) are maximum entangled states of two-photon
and three-photon, respectively. In transmission process, they
should become non-maximally entangled states, they are
\begin{eqnarray}
|\psi\rangle=c_1|HH\rangle+c_2|VV\rangle,
\end{eqnarray}
\begin{eqnarray}
|GHZ\rangle=c_3|HHH\rangle+c_4|VVV\rangle,
\end{eqnarray}
where $c_1$, $c_2$, $c_3$ and $c_4$ are real or plural
coefficients, and satisfying with the normalization conditions
$|c_1|^2+|c_2|^2=1$ and $|c_3|^2+|c_4|^2=1$. The quantum entangled
degree of entangled states (7) and (8) are
\begin{eqnarray}
E=-(|c_1|^2\log_2|c_1|^2+|c_2|^2\log_2|c_2|^2).
\end{eqnarray}
\begin{eqnarray}
E=-(|c_3|^2\log_2|c_3|^2+|c_4|^2\log_2|c_4|^2).
\end{eqnarray}
With Eqs. (9) and (10), the quantum entangled degree of the
two-photon and three-photon entangled states (5) and (6) are $1$,
i.e., maximum entanglement. For the entangled states (7) and (8),
their quantum entangled degrees depend on the coefficients $c_1$,
$c_2$ and  $c_3$, $c_4$. When the two and three entangled photon
propagate in space, the coefficients $c_1$, $c_2$ and  $c_3$,
$c_4$ should be changed, and the quantum entangled degree should
be decreased and even approach zero, the quantum entanglement of
two and three photon shall be disappeared. All the quantum
information processes, e.g., quantum communication, quantum
computation and so on can not proceed. The attenuation of quantum
entangled degree is unavoidable, but we can magnify the quantum
entangled degree in the transmission process of entangled photon,
which can be achieved by one-dimensional photonic crystal.

The Eq. (4) gives the relation between the output and the input
electric field intensity in one-dimensional photonic crystal, it
is
\begin{eqnarray}
E_{out}=tE_{in},
\end{eqnarray}
where $t$ is the transmission coefficient, $E_{out}$ and $E_{in}$
are output and input electric field intensity respectively. In
Eqs. (7) and (8), when the two and three entangled photon entering
one-dimensional photonic crystal, the output electric field
intensity of the horizontal ($TM$) state $|H\rangle$ and the
vertical ($TE$) state $|V\rangle$ are
\begin{eqnarray}
E_{out}^H=t_ME_{in}^H,
\end{eqnarray}
and
\begin{eqnarray}
E_{out}^E=t_EE_{in}^E,
\end{eqnarray}
where $t_M$ ($t_E$) is the transmission coefficient of horizontal
(vertical) state $|H\rangle$ ($|V\rangle$), $E_{out}^H$
($E_{out}^E$) and $E_{in}^H$ ($E_{in}^E$)are the output and input
electric field intensity of horizontal (vertical) state
$|H\rangle$ ($|V\rangle$) in one-dimensional photonic crystal.

The entangled states (7) and (8) are as input entangled states
\begin{eqnarray}
|\psi\rangle_{in}=c_1|HH\rangle+c_2|VV\rangle,
\end{eqnarray}
\begin{eqnarray}
|GHZ\rangle_{in}=c_3|HHH\rangle+c_4|VVV\rangle,
\end{eqnarray}
passing through one-dimensional photonic crystal, their output
entangled states become
\begin{eqnarray}
|\psi\rangle_{out}=c_1t_M^2|HH\rangle+c_2t_E^2|VV\rangle,
\end{eqnarray}
\begin{eqnarray}
|GHZ\rangle_{out}=c_3t_M^\frac{3}{2}|HHH\rangle+c_4t_E^\frac{3}{2}|VVV\rangle,
\end{eqnarray}
\begin{figure}[tbp]
\includegraphics[width=8cm, height=2cm]{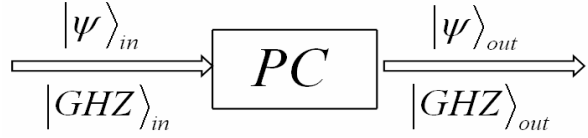}
\caption{The input entangled states $|\psi\rangle_{in}$ and
$|GHZ\rangle_{in}$ pass through the photonic crystal $PC$ become
the output entangled states $|\psi\rangle_{out}$ and
$|GHZ\rangle_{out}$.}
\end{figure}
the input entangled states and output entangled states are shown
in FIG. 1, the $PC$ express the one-dimensional photonic crystal.

Eqs. (16) and (17) normalization form are
\begin{eqnarray}
|\psi\rangle_{out}=d_1|HH\rangle+d_2|VV\rangle,
\end{eqnarray}
\begin{eqnarray}
|GHZ\rangle_{out}=d_3|HHH\rangle+d_4|VVV\rangle,
\end{eqnarray}
where normalization constants are
\begin{eqnarray}
d_1=\frac{c_1t_M^2}{\sqrt{|c_1t_M^2|^2+|c_2t_E^2|^2}}, \hspace
{0.1in} d_2=\frac{c_2t_E^2}{\sqrt{|c_1t_M^2|^2+|c_2t_E^2|^2}},
\end{eqnarray}
\begin{eqnarray}
d_3=\frac{c_3t_M^\frac{3}{2}}{\sqrt{|c_3t_M^\frac{3}{2}|^2+|c_4t_E^\frac{3}{2}|^2}},
\hspace {0.1in}
d_4=\frac{c_4t_E^\frac{3}{2}}{\sqrt{|c_3t_M^\frac{3}{2}|^2+|c_4t_E^\frac{3}{2}|^2}},
\end{eqnarray}
the quantum entangled degree of output entangled states (18) and
(19) are
\begin{eqnarray}
E=-(|d_1|^2\log_2|d_1|^2+|d_2|^2\log_2|d_2|^2).
\end{eqnarray}
\begin{eqnarray}
E=-(|d_3|^2\log_2|d_3|^2+|d_4|^2\log_2|d_4|^2).
\end{eqnarray}
where
\begin{eqnarray}
|d_1|^2=\frac{|c_1|^2T_M^2}{\sqrt{|c_1|^2T_M^2+|c_2|^2T_E^2}},
\hspace {0.1in}
|d_2|^2=\frac{|c_2|^2T_E^2}{\sqrt{|c_1|^2T_M^2+|c_2|^2T_E^2}},
\end{eqnarray}
\begin{eqnarray}
|d_3|^2=\frac{|c_3|^2T_M^{\frac{3}{2}}}
{\sqrt{|c_3|^2T_M^{\frac{3}{2}}+|c_4|^2T_E^{\frac{3}{2}}}},
\hspace {0.1in}
|d_4|^2=\frac{|c_4|^2T_E^{\frac{3}{2}}}{\sqrt{|c_3|^2T_M^{\frac{3}{2}}+|c_4|^2T_E^{\frac{3}{2}}}}.
\end{eqnarray}
From Eqs. (24) and (25) we can find the quantum entangled degree
of output entangled states are related to the transmissivity $T_M$
and $T_E$ of one-dimensional photonic crystal. In one-dimensional
photonic crystal, their transmissivity are $0\leq T_M\leq1$ and
$0\leq T_E\leq1$. In the following calculation, we shall find the
quantum entangled degree of output entangled states can be
magnified by the given transmissivity $T_M$ and $T_E$ values.
\begin{figure}[tbp]
\includegraphics[width=12cm, height=6cm]{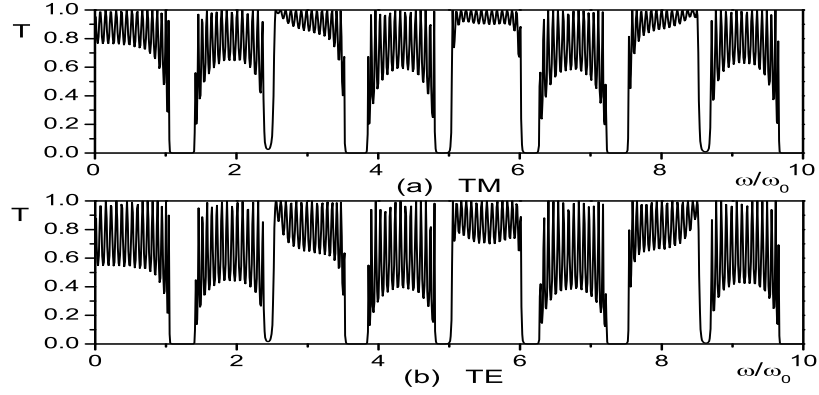}
\caption{The one-dimensional photonic crystal transmissivity $T_M$
and $T_E$ for $TM$ and $TE$ waves.}
\end{figure}
\begin{figure}[tbp]
\includegraphics[width=12cm, height=6cm]{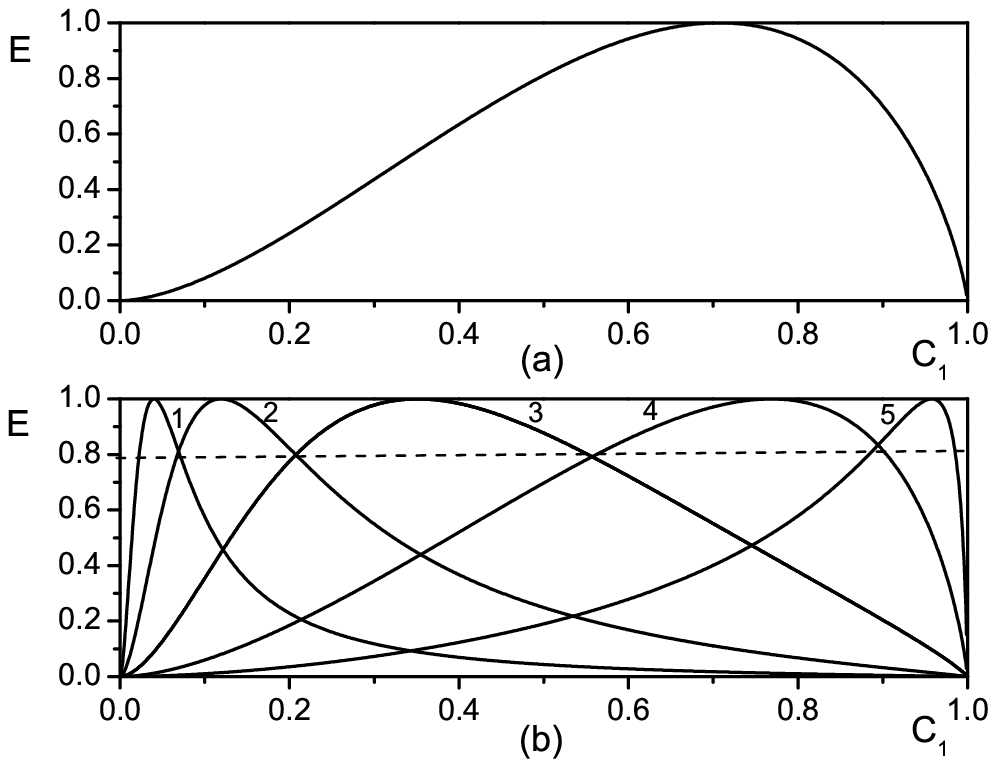}
\caption{The quantum entangled degree of two-photon input and
output entangled states.}
\end{figure}
\begin{figure}[tbp]
\includegraphics[width=12cm, height=6cm]{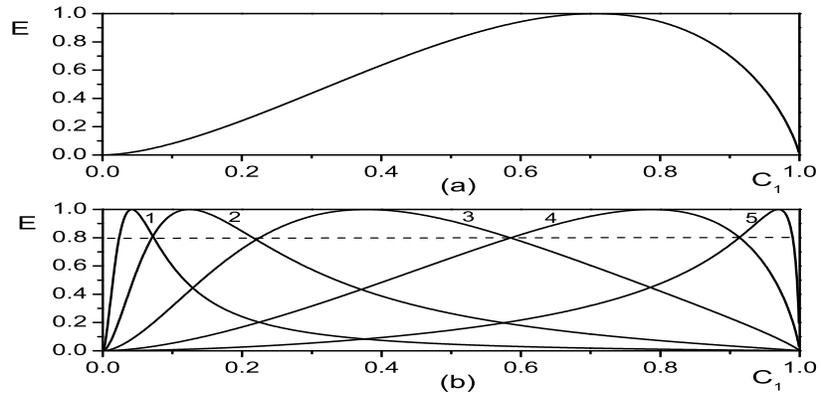}
\caption{The quantum entangled degree of three-photon input and
output entangled states.}
\end{figure}

\vskip 8pt {\bf 4. Numerical result} \vskip 8pt

In this section, we report our numerical results of the quantum
entangled degrees for the two-photon and three-photon input and
output entangled states. Firstly, we should calculate the
transmissivity $T_M$ and $T_E$ for the $TM$ and $TE$ waves. The
main parameters are: The center angle wavelength
$\omega_0=1.22\cdot10^{15}Hz$, the medium $A$ refractive indices
$n_a=1.68$, thickness $a=108nm$, the medium $B$ refractive indices
$n_b=2.56$, thickness $b=198nm$, the structure is $(AB)^{16}$ and
the incident angle $\theta_0=\pi/6$. By Eq. (4), we can calculate
the transmissivity of $T_M$ and $T_E$, which are shown in FIG. 2.
FIG. 2 (a) and (b) are the transmissivity of $T_M$ and $T_E$,
respectively. We can find their transmissivity are $0\leq
T_M\leq1$ and $0\leq T_E\leq1$. We shall use the $T_M$ and $T_E$
values in calculating the quantum entangled degrees. Secondly, by
Eq. (9), we calculate the quantum entangled degrees of the
two-photon input entangled state (14), which is shown in FIG. 3
(a), it gives the relation between coefficient $c_1$ and the
entangled degrees $E$ of input entangled state. By Eqs. (22) and
(24), we calculate the quantum entangled degrees of the two-photon
output entangled state (16), which is shown in FIG. 3 (b), there
are five output quantum entangled degrees curves. The curves $1$,
$2$, $3$, $4$ and $5$ are obtained by taking the transmissivity
$T_E=0.04$ and $T_M=1.0$, $T_E=0.12$ and $T_M=1.0$, $T_E=0.30$ and
$T_M=0.80$, $T_E=0.60$ and $T_M=0.50$, $T_E=1.0$ and $T_M=0.30$
respectively. The five groups different transmissivity can be
easily realized by five different structure one-dimensional
photonic crystals. From FIG. 3 (a), we can find the entangled
degrees $E$ relaxedly increases from $0$ to $1$ when the
coefficient $c_1$ is in the range of $0\sim\frac{1}{\sqrt{2}}$,
and the entangled degrees $E$ relaxedly decreases from $1$ to $0$
when the coefficient $c_1$ is in the range of
$\frac{1}{\sqrt{2}}\sim 1$, i.e., the input quantum entangled
degrees of two-photon tend to $1$ in the vicinity of
$c_1=\frac{1}{\sqrt{2}}$, and it is relatively low in the other
area of $c_1$. Then, the input entangled state of two-photon can
not proceed long range propagation. In FIG. 3 (b), we can find the
output quantum entangled degrees $E$, constituted by five output
quantum entangled degrees curves, the quantum entangled degrees
$E$ is in the range of $0.8\sim 1$ when $c_1=0.02\sim 0.99$, which
has been magnified. When the output quantum entangled degrees can
be magnified continuously, the output entangled state of
two-photon should be achieved long range propagation. Finally, for
the three-photon input and output entangled states (15) and (19),
we have similarly obtained the input and output quantum entangled
degrees, which are shown in FIG. 4 (a) and (b), and the results
are the same as the two-photon input and output quantum entangled
degrees.

\vskip 10pt {\bf 5. Conclusion} \vskip 10pt

In summary, we have designed a new type quantum entangled degrees
amplifier by one-dimensional photonic crystal, which is similar as
the relay station of classical electromagnetic communication. By
calculation, we find when the entangled states of two-photon and
three-photon pass through photonic crystal, their quantum
entanglement degrees can be magnified, which make the entanglement
states of two-photon, three-photon and multi-photon can be long
range propagation and the quantum communication can be really
realized.

\vskip 12pt {\bf 6.  Acknowledgment} \vskip 12pt

This work was supported by the National Natural Science Foundation
of China (no.61275047), the Research Project of Chinese Ministry
of Education (no.213009A) and Scientific and Technological
Development Foundation of Jilin Province (no.20130101031JC).

\end{document}